\documentclass[11pt,nohyper,a4paper]{JHEP3}
\usepackage{amssymb,euscript,bbold}
\newcommand{\rep}[1]{\mathbf{#1}}

\newcommand{\be}{\begin{equation}}
\newcommand{\ee}{\end{equation}}
\newcommand{\bea}{\begin{eqnarray}}
\newcommand{\eea}{\end{eqnarray}}
\newcommand{\ba}{\begin{array}}
\newcommand{\p}[1]{(\ref{#1})}
\newcommand{\ea}{\end{array}}

\def\bbox{{\,\lower0.9pt\vbox{\hrule \hbox{\vrule height 0.2 cm
\hskip 0.2 cm \vrule height 0.2 cm}\hrule}\,}}
\newcommand{\dsl}{\pa \kern-0.5em /}

\newcommand{\nn}{\nonumber \\}

\def\CL{{\cal L}}                       

\def\CD{{\cal D}}

\def\cF{{\cal F}}

\def\ds{\raise.15ex\hbox{/}\kern-.57em\partial}
\def\Ds{\,\raise.15ex\hbox{/}\mkern-13.5mu D}

\renewcommand{\a}{\alpha}
\renewcommand{\b}{\beta}

\newcommand{\e}{\epsilon}
\newcommand{\K}{{{K}}}

\newcommand{\aomega}{\alpha}

\font\mybb=msbm10 at 10pt
\def\bb#1{\hbox{\mybb#1}}

\def\bR {\bb{R}}

\def\a{\alpha}\def\b{\beta}
\def\s{\sigma}

\title{The Geometry of D=11 Null Killing
Spinors}

\author{Jerome P. Gauntlett\\
Perimeter Institute for Theoretical Physics\\
Waterloo, ON, N2J 2W9, Canada \thanks{
On leave from: Blackett Laboratory, Imperial College,
London, SW7 2BZ, U.K.}\\ E-mail:
\email{jgauntlett@perimeterinstitute.ca}}

\author{Jan B. Gutowski\\ Mathematical Institute\\
Oxford, OX1 3LB, UK\\ E-mail:
\email{gutowski@maths.ox.ac.uk}}

\author{Stathis Pakis\\Blackett Laboratory, Imperial College\\
London, SW7 2BZ, U.K   \\ E-mail:
\email{s.pakis@imperial.ac.uk}}

\abstract{
We determine the necessary and sufficient conditions on the metric
and the four-form for the most general bosonic supersymmetric
configurations of D=11 supergravity which admit a null Killing
spinor i.e. a Killing spinor which can be used to construct a null
Killing vector. This class covers all supersymmetric
time-dependent configurations and completes the classification of
the most general supersymmetric configurations initiated in
\cite{gauntpakis}.}


\preprint{Imperial/TP/03-04/3}

\begin{document}

\setcounter{equation}{0}

\section{Introduction}

Supersymmetric bosonic solutions of supergravity theories have
been important in many developments in string/M-theory. Recently
there has been significant progress in determining the most
general kinds of geometries that underly all such solutions of a
particular theory \cite{gauntpakis, mostqmw, gauntgut,
martgutreal:2003,Caldarelli:2003pb}. In addition to providing a
deeper understanding of known classes of solutions, this analysis
is useful in precisely characterising geometries of interest when
explicit solutions are difficult to come by. Thus, for example, a
uniqueness theorem for a class of supersymmetric black holes was
found in \cite{harvey}. This analysis also provides new techniques
for constructing explicit solutions that have been hitherto missed
by other approaches based on guessing ansatz\"e.

The basic idea is to translate the condition for supersymmetry,
the existence of a Killing spinor, into differential conditions on
the tensors that can be constructed as bi-linears from the Killing
spinor. This approach was first employed by Tod some time ago
\cite{tod,Tod:jf}, following \cite{gibbhull}, to analyse N=2
supergravity in D=4. For the case of the minimal supergravity theory,
using features specific to D=4, it was possible to explicitly
construct the most general supersymmetric solutions. In higher
dimensions, it is not possible to find all of the solutions in
closed form but nevertheless, a precise description of the
geometry can be made.

One of the most interesting supergravity theories to study is D=11
supergravity as it describes the low-energy limit of M-theory. In
\cite{gauntpakis}, a programme for classifying all supersymmetric
solutions of D=11 supergravity was outlined based on the above
strategy. A key observation of \cite{gauntpakis} is that in
organising the calculations it is very useful to note that the
Killing spinors define a privileged $G$-structure. That is, in
this case, a reduction of the $Spin(10,1)$ frame bundle to a
$G$-sub-bundle (for some general mathematical discussion on
$G$-structures see e.g. \cite{sal1}). The utility of
$G$-structures in analysing restricted classes of supergravity
solutions has also been shown in \cite{friv}-\cite{Behrndt:2003ih}
(see also \cite{Warner,Warner2}).

One general feature of the results of \cite{gauntpakis} is that
generalised calibrations \cite{Gutowski:1999iu,gpt,blw} naturally
emerge from the conditions for supersymmetry. This was noticed
earlier for a restricted class of configurations of D=10
supergravity in \cite{Gauntlett:2001ur}, and is related to the
fact that supersymmetric geometries with non-vanishing fluxes
arise when branes wrap calibrated cycles in special holonomy
manifolds, after taking the back reaction into account. It was
shown in \cite{HPS} that the connections with generalised
calibrations found in \cite{gauntpakis} lead to a simple proposal
for the topological charges appearing in the supersymmetry
algebras of membranes and fivebranes propagating in general
supersymmetric backgrounds.

The most general supersymmetric solutions of D=11 supergravity
preserve at least one supersymmetry and hence admit at least one
Killing spinor. However, there are two kinds of spinors of
Spin(10,1) which are distinguished by whether the vector that can
be constructed as a bi-linear from the spinor is time-like or null
\cite{bryant}. Moreover, the vector constructed from the Killing
spinor is necessarily Killing. Consequently there are two kinds of
supersymmetric solutions of D=11 supergravity: those admitting a
 ``time-like Killing spinor'', which can be used
to construct a time-like Killing vector, and those admitting a
``null Killing spinor'', which can be used to  construct a null
Killing vector\footnote{Note that geometries preserving more than
one supersymmetry can be in both classes. Also note, for example,
that it is possible for a geometry preserving just one
supersymmetry to be in the null class and also admit a time-like
Killing vector, but the time-like Killing vector will not be built
from the Killing spinor.}.

Supersymmetric solutions with a single time-like Killing spinor
were analysed in detail in \cite{gauntpakis}. It was shown that
the Killing spinor defines a preferred $SU(5)$-structure and this
was used to determine the most general local form of the metric
and the four-form field strength. The focus of this paper will be
to perform a similar analysis for supersymmetric solutions
admitting a null Killing-spinor. These solutions have a
$(Spin(7)\ltimes\bR^8)\times \bR$ structure, which can be used to
determine the general local form of the solutions. It is worth
emphasising that since all of the solutions in the time-like class
are stationary, in this sense any time-dependent supersymmetric
solutions are necessarily in the null class.

The results of this paper and \cite{gauntpakis} therefore provide
a classification of the most general supersymmetric solutions of
D=11 supergravity, preserving at least 1/32 supersymmetry. This
classification can in principle be refined \cite{gauntpakis} by
analysing the additional conditions placed on the geometries
preserving more than one supersymmetry. For example, if the
geometry preserves two supersymmetries, the two Killing spinors
could both be null, both time-like or be one of each. These
solutions will thus be special cases of the geometries presented
here, or in \cite{gauntpakis} (or both), satisfying extra
constraints. Similarly, solutions preserving more supersymmetries
will be further restricted. It will be very interesting to pursue
this further. It is noteworthy that the classification of
maximally supersymmetric configurations has already been carried
out in \cite{figpap} using methods specific to this case.

The plan of this paper is as follows. In section 2 we evaluate the
algebraic and differential identities which the various bi-linears
constructed from a null Killing spinor must satisfy. In section 3
we use these expressions to constrain the eleven-dimensional
geometry and to fix almost all of the components of the four-form.
We also demonstrate that these necessary conditions are in fact
sufficient by demonstrating that the configurations always admit a
Killing spinor. For the convenience of the reader we have
summarised the main results of this section in section 3.3. In
section 4 we introduce local co-ordinates on the
eleven-dimensional manifold in which the constraints take a rather
simple form. In particular, we show that the $Spin(7)$ invariant
4-form must be conformally anti-self-dual. We again summarise the
main result of this section in a separate sub-section. We
demonstrate in section 5 how the resolved membrane solution of
\cite{Cvetic:2000mh}, the 1/32 supersymmetric membrane/wave solution
of \cite{Gomis:2003zw}, and the basic fivebrane solution can be
obtained from our construction. This provides non-trivial checks
on our calculations as well as providing some intuition into the
kinds of $(Spin(7)\ltimes\bR^8)\times \bR$ structures allowed by
supersymmetry. In addition, the formalism allows us to generalise
the resolved membrane solution by the addition of a gravitational
wave thus combining the solutions of \cite{Cvetic:2000mh,Gomis:2003zw}.
In section 6 we analyse the special case when the four-form
vanishes, recovering the general local-form for the solution found
in \cite{bryant}. Section 7 briefly concludes. The paper finishes
with several Appendices containing some useful technical
information.

\section{Killing spinors and differential forms}
The bosonic fields of D=11 supergravity consist of a metric, $g$,
and a three-form potential $C$ with four-form field strength
$F=dC$. The action for the bosonic fields is given by \bea
S=\frac{1}{2\kappa^2}\int d^{11} x {\sqrt{-g}}R
-\frac{1}{2}F\wedge *F - \frac{1}{6}C\wedge F\wedge F \ . \eea The
equations of motion\footnote{Note that in  M-theory, the
field-equation for the four-form receives higher order
gravitational corrections given, in our conventions, by equation
(2.4) in \cite{gauntpakis}. Since most of our analysis only
concerns the Killing spinor equation given here in \p{killing},
including this correction at the level of the gauge equations of motion is
straightforward.}  and the Bianchi
identity are thus given by \bea \label{eqn:einst}
R_{\mu\nu}-\frac{1}{12}(F_{\mu
\s_1\s_2\s_3}F{_{\nu}}{^{\s_1\s_2\s_3}}-
\frac{1}{12}g_{\mu\nu}F^2)&=&0\nn d*F+\frac{1}{2}F\wedge F&=&0\nn
dF&=&0 \ , \eea where $F^2=F_{\mu_1\mu_2\mu_3\mu_4}
F^{\mu_1\mu_2\mu_3\mu_4}$. A solution of these equations preserves
at least one supersymmetry if it admits at least one Killing
spinor, $\epsilon$, which solves
\begin{equation}\label{killing}
\nabla_\mu\e+\frac{1}{288}[\Gamma{_\mu}{^{\nu_1\nu_2\nu_3\nu_4}}
-8\delta{_\mu^{\nu_1}}\Gamma^{\nu_2\nu_3\nu_4}]F_{\nu_1\nu_2\nu_3\nu_4}
\e=0 \ .
\end{equation}
Our conventions are outlined in appendix A.

Consider a configuration $(g,F)$ that admits a single Killing
spinor $\e$. We can then define the following  one-, two- and
five-forms: \bea\label{defforms} \K_\mu&=&\bar\e\Gamma_\mu\e \nn
\Omega_{\mu_1\mu_2}&=&\bar\e\Gamma_{\mu_1\mu_2}\e\nn
\Sigma_{\mu_1\mu_2\mu_3\mu_4\mu_5}&=&\bar\e\Gamma_{\mu_1\mu_2\mu_3
\mu_4\mu_5}\e \ . \eea
Note that in the above construction we take $\e$ to be a
commuting spinor. Of course the
supersymmetry parameter is an anticommuting spinor but since we
are interested
in purely bosonic supersymmetric configurations the only relevant
supersymmetry
variation is that of the gravitino which yields the Killing spinor
equation. This
is linear in the spinor and hence the existence of a
commuting Killing spinor
is equivalent to the preservation of a supersymmetry. As noted
in \cite{gauntpakis},
using an argument presented in \cite{harvey}, we can assume
without loss of
generality that  $\e$ is nowhere vanishing. If there is more than
one linearly independent Killing spinor, then additional forms
including scalars, three and four-forms can also be defined.
However, here we shall only consider the most general case of a
single Killing spinor.

These differential forms are not all independent. They satisfy
certain algebraic relations which are consequences of the
underlying Clifford algebra. The traditional way of obtaining
these is by repeated use of Fierz identities and some were
presented in \cite{gauntpakis}. Alternatively we can use the fact
that the forms, or equivalently the Killing spinor, give rise to
privileged $G$-structures with $G\subset Spin(10,1)$. Indeed it
was argued in \cite{gauntpakis}, using results of \cite{bryant},
that there are two possibilities. If $K$ is null everywhere then
the forms give rise to a globally defined
$(Spin(7)\ltimes\bR^8)\times \bR$ structure. If $K$ is timelike at
a point, on the other hand, then it is timelike in a neighbourhood
of this point and the forms then define a privileged $SU(5)$
structure in this neighbourhood. It is not possible to have a
spacelike $K$.

The necessary and sufficient conditions for a configuration
$(g,F)$ to admit time-like Killing spinors were analysed in
\cite{gauntpakis}. Here we will focus on the null case. It is
therefore convenient to work in a null basis \be\label{met}
ds^2=2e^+e^- + e^i e^i  +e^9e^9 \ee with $i=1,\dots,8$ and \be
K=e^+ \ . \ee We choose an orientation such that
$\e_{+-123456789}=-1$. The most convenient way to determine the
forms $\Omega$ and $\Sigma$ for null spinors is to construct them
from a specific null spinor. Such a spinor can be fixed by
demanding that it satisfies the following
projections\footnote{From a physical point of view these
projections are equivalent to those arising when a fivebrane wraps
a Cayley four-cycle \cite{glw}. Corresponding supergravity
solutions were presented in \cite{Gauntlett:2000ng}.}
\bea\label{projs}
\Gamma_{1234}\e=\Gamma_{3456}\e=\Gamma_{5678}\e=\Gamma_{1357}\e=-\e\nn
\Gamma^+\e=0 \ . \eea Note that these conditions automatically
imply that $\Gamma^9\e=\e$. It is then straightforward to deduce
that \bea \Omega&=&e^+\wedge e^9\nn \Sigma&=&e^+\wedge \phi \eea
where $\phi=\frac{1}{4!}\phi_{i_1i_2i_3i_4}e^{i_1i_2i_3i_4}$ is
the Spin(7) invariant four-form whose only non-vanishing
components are given by \bea \label{phidef}
   -\phi &= e^{1234}+e^{1256}+e^{1278}+e^{3456}+e^{3478}+e^{5678} \nn
      &\qquad + e^{1357}-e^{1368}-e^{1458}-e^{1467}-e^{2358}
          -e^{2367}-e^{2457}+e^{2468} \ .
\eea Observe that the action of $(Spin(7) \ltimes \bR^8) \ltimes
\bR$ on the basis 1-forms is given by \bea \label{eqn:basisrot}
e^+ &\rightarrow& (e^+)' = e^+ \nn e^-  &\rightarrow& (e^-)'= e^-
-{1 \over 2} (\alpha^2+p_i p^i) e^+ -
 \alpha e^9 -Q_{i j}p^i e^j
\nn e^9 &\rightarrow& (e^9)'= e^9+ \alpha e^+ \nn e^i
&\rightarrow& (e^i)' =Q^i{}_j e^j + p^i e^+ \ , \eea where $Q \in
Spin(7)$, and $p_i = \delta_{ij}p^j$; in particular, we see that
these transformations not only preserve the metric but also $K,
\Omega$ and $\Sigma$.

In appendix B we present an alternative derivation of these
results using Fierz identities; in particular this provides an
independent check of some of the results of \cite{bryant}.

The covariant derivatives of the differential forms were
calculated in \cite{gauntpakis}. The result for both the timelike
and the null case is: \bea \label{cova} \nabla_\mu K_\nu &=&{1
\over 6} \Omega^{\sigma_1 \sigma_2}F_{\sigma_1 \sigma_2 \mu \nu}
+{1 \over 6!} \Sigma^{\sigma_1 \sigma_2 \sigma_3 \sigma_4
\sigma_5} *F_{\sigma_1 \sigma_2 \sigma_3 \sigma_4 \sigma_5 \mu
\nu} \eea \bea \label{covb} \nabla_\mu \Omega_{\nu_1 \nu_2} &=& {1
\over 3.4!}g_{\mu [\nu_1} \Sigma_{\nu_2]}{}^{\sigma_1 \sigma_2
\sigma_3 \sigma_4} F_{\sigma_1 \sigma_2 \sigma_3 \sigma_4} +{1
\over 3.3!} \Sigma_{\nu_1 \nu_2}{}^{\sigma_1 \sigma_2 \sigma_3}
F_{\mu \sigma_1 \sigma_2 \sigma_3} \nn &-&{1 \over 3.3!}
\Sigma_{\mu [\nu_1}{}^{\sigma_1 \sigma_2 \sigma_3}F_{\nu_2]
\sigma_1 \sigma_2 \sigma_3} +{1 \over 3} K^\sigma F_{\sigma \mu
\nu_1 \nu_2} \eea \bea \label{covc} \nabla_\mu \Sigma_{\nu_1 \nu_2
\nu_3 \nu_4 \nu_5} &=& {1 \over 6} K^\sigma *F_{\sigma \mu \nu_1
\nu_2 \nu_3 \nu_4 \nu_5}-{10 \over 3}F_{\mu [\nu_1 \nu_2 \nu_3}
\Omega_{\nu_4 \nu_5]} -{5 \over 6}F_{[\nu_1 \nu_2 \nu_3
\nu_4}\Omega_{\nu_5] \mu} \nn &-&{10 \over 3} g_{\mu [\nu_1}
\Omega_{\nu_2}{}^\sigma F_{|\sigma| \nu_3 \nu_4 \nu_5]} +{5 \over
6} F_{\mu [\nu_1 |\sigma_1 \sigma_2|} *\Sigma^{\sigma_1
\sigma_2}{}_{\nu_2 \nu_3 \nu_4 \nu_5]} \nn &+&{5 \over 6}
F_{[\nu_1 \nu_2 |\sigma_1 \sigma_2|} *\Sigma^{\sigma_1
\sigma_2}{}_{\nu_3 \nu_4 \nu_5] \mu} -{5 \over 9} g_{\mu
[\nu_1}F_{\nu_2 |\sigma_1 \sigma_2 \sigma_3|} * \Sigma^{\sigma_1
\sigma_2 \sigma_3}{}_{\nu_3 \nu_4 \nu_5]} \ . \nn \eea
 The exterior derivatives of the forms are thus given by
\bea\label{dk}
d\K&=&\frac{2}{3}i_{\Omega}F+\frac{1}{3}i_{\Sigma}*F\\
\label{dktwo}
d\Omega&=&i_{K}F\\
\label{dkthree} d\Sigma&=&i_{K}* F -\Omega\wedge F \eea where e.g.
$(i_\Omega F)_{\mu\nu}=(1/2!)\Omega^{\rho_1\rho_2}
{}F_{\rho_1\rho_2\mu\nu}$.

{}From the first equation in \p{cova} we can immediately deduce
that $K$ is a Killing vector. Moreover, using the Bianchi
identity, it is simple to show that \be\label{ksym} \CL_{K} F=0 \
.\ee Thus any geometry $(g,F)$ admitting a Killing spinor
possesses a symmetry generated by $K$. In addition, the
Lie-derivatives of $\Omega$ and $\Sigma$ with respect to $K$ also
vanish: \bea \label{liederstr} \CL_K \Omega&=&0\nn \CL_K
\Sigma&=&0 \ . \eea

\section{The geometry of null Killing spinors}
Our aim is to extract from the differential conditions \p{cova},
\p{covb} and \p{covc} the necessary and sufficient conditions on
the geometry and the four-form field strength in order that they
admit null Killing spinors. In the next two subsections we derive
the necessary conditions and then show that they are sufficient.
In the third subsection we have summarised the results. The final
brief subsection states the extra conditions required in order
that the configuration also solves the equations of motion.

\subsection{Necessary conditions}
We will need to decompose various forms carrying totally
anti-symmetric $SO(8)$ representations into Spin(7) reps. If we
denote by $\Lambda^p$ the space of $p$-forms constructed from
$e^i$ only, we have the  following decompositions: \bea
\Lambda^1:&\rep{8}\to& \rep{8}\nn \Lambda^2:&\rep{28}\to& \rep{21}
+\rep{7}\nn \Lambda^3:&\rep{56}\to& \rep{48} +\rep{8}\nn
\Lambda^4:&\rep{70}\to& \rep{35} + \rep{1} +\rep{7}+\rep{27} \ .
 \eea
For two-forms the projections can be written explicitly as \bea
 (P^{\rep{21}} \aomega)_{i_1 i_2} &=&\frac{3}{4}(\aomega_{i_1
i_2}+\frac{1}{6}\phi_{i_1 i_2}{^{j_1 j_2}}
 \aomega_{j_1 j_2})\nn
(P^{\rep{7}} \aomega)_{i_1 i_2} &=&\frac{1}{4}(\aomega_{i_1
i_2}-\frac{1}{2}\phi_{i_1 i_2}{^{j_1 j_2}}
 \aomega_{j_1 j_2}) \ .
\eea For three-forms we have,
\begin{eqnarray}
(P^{\rep{48}} \aomega)_{i_1 i_2 i_3}&=&\frac{6}{7}(\aomega_{i_1
i_2 i_3} +\frac{1}{4}\phi^{j_1j_2}{}_{[i_1 i_2} \aomega_{i_3]
j_1j_2})\nn (P^{\rep{8}} \aomega)_{i_1 i_2
i_3}&=&\frac{1}{7}(\aomega_{i_1 i_2 i_3}
-\frac{3}{2}\phi^{j_1j_2}{}_{[i_1 i_2} \aomega_{i_3] j_1j_2}) \ .
\end{eqnarray}
For four-forms the $\rep{35}$ is the anti-self-dual piece, while
\bea\label{projfour} (P^{\rep{1}} \aomega)_{i_1 i_2 i_3 i_4} &=&{1
\over 336} \phi_{i_1 i_2 i_3 i_4} \phi^{j_1j_2j_3j_4}
\aomega_{j_1j_2j_3j_4} \nn (P^{\rep{7}} \aomega)_{i_1 i_2 i_3 i_4}
&=& {1 \over 8}
 \aomega_{i_1 i_2 i_3 i_4}
+{1 \over 224} \phi_{i_1 i_2 i_3 i_4} \phi^{j_1j_2j_3j_4}
\aomega_{j_1j_2j_3j_4} \nn &-&{3 \over 224} \phi^{j_1j_2}{}_{[i_1
i_2} \phi_{i_3 i_4]}{}^{j_3j_4} \aomega_{j_1j_2j_3j_4} +{5 \over
168} \phi^{j_1}{}_{[i_1 i_2 i_3} \phi_{i_4]}{}^{j_2j_3j_4}
\aomega_{j_1j_2j_3j_4} \nn (P^{\rep{27}} \aomega)_{i_1 i_2 i_3
i_4} &=& {3 \over 8} \aomega_{i_1 i_2 i_3 i_4} -{1 \over 224}
\phi_{i_1 i_2 i_3 i_4} \phi^{j_1j_2j_3j_4} \aomega_{j_1j_2j_3j_4}
\nn &+&{15 \over 224} \phi^{j_1j_2}{}_{[i_1 i_2} \phi_{i_3
i_4]}{}^{j_3j_4} \aomega_{j_1j_2j_3j_4} +{1 \over 56}
\phi^{j_1}{}_{[i_1 i_2 i_3} \phi_{i_4]}{}^{j_2j_3j_4}
\aomega_{j_1j_2j_3j_4} \ . \eea The identities satisfied by $\phi$
that we used to construct these projections, as well as various
identities satisfied by the forms in different representations are
presented in appendix B. Note that as far as we know, the
projections for the four-forms are new.

By analyzing the expressions  \p{dk}, \p{dktwo} for $dK$ and
$d\Omega$ we immediately find that some components of the flux
must vanish: \bea F_{-i_1i_2i_3}&=&0\nn F^\rep{7}_{-i_1i_29}&=&0 \
. \eea In addition we get \bea \label{eqn:spncona}
dK=de^+&=&(-\frac{2}{3}F_{+-i9}+\frac{1}{3}\frac{1}{3!}
\phi_i{}^{j_1j_2j_3} F_{j_1j_2j_39})e^{+i}\nn
&+&\frac{1}{3}(\frac{1}{4!}\phi^{i_1i_2i_3i_4}F_{i_1i_2i_3i_4})e^{+9}
+\frac{1}{2}(F_{-i_1i_29})e^{i_1i_2} \ . \eea Note that $K$
satisfies \be
 K\wedge
dK=\frac{1}{2}(F_{-i_1i_29})e^{+i_1i_2} \ee which implies that $K$
is not hyper-surface orthogonal in general\footnote{In the
most general supersymmetric geometries of D=6 minimal supergravity
there is always a null Killing vector which is also not
hyper-surface orthogonal in general \cite{martgutreal:2003}.}.

To proceed it is useful to write the constraints on $F$ in terms
of the spin connection $\omega$ defined by
\begin{eqnarray}
(\nabla_{\a} e^{\b})_\lambda \equiv -\omega_\a{}^\b{}_\lambda \ .
\end{eqnarray}
In particular, as $K$ is Killing, we have \be \omega_{(\alpha
\beta) -}=0 \ee and
 ({\ref{eqn:spncona}}) can be rewritten as
\bea \omega_{+i-} &=& -{1 \over 3} F_{+-i9}+{1 \over 36}
\phi_i{}^{j_1 j_2 j_3} F^\rep{8}_{j_1 j_2 j_3 9} \nn \omega_{+9-}
&=& {1 \over 144} \phi^{i_1 i_2 i_3 i_4} F^\rep{1}_{i_1 i_2 i_3
i_4} \nn \omega_{i_1i_2-} &=& {1 \over 2} F_{-9i_1i_2} \nn
\omega_{i9-} &=& 0 \ . \eea Next we examine ({\ref{covb}}). This
implies the following additional relationships between the spin
connection and the gauge field strength: \bea
(\nabla_{+}\Omega)_{+i}\Rightarrow \qquad \omega_{+9i} &=&
-\frac{1}{12}\phi_i{^{j_1 j_2 j_3}}F^{\rep{8}}_{+ j_1 j_2 j_3} \nn
(\nabla_{+}\Omega)_{i_1i_2}\Rightarrow \qquad
F^{\rep{7}}_{+-i_1i_2} &=& {1 \over 24} \phi^{j_1 j_2
j_3}{}_{[i_1} F^\rep{7} _{i_2] j_1 j_2 j_3} \nn
(\nabla_{-}\Omega)_{+i}\Rightarrow \qquad \omega_{-9i} &=& 0 \nn
(\nabla_{i_1}\Omega)_{+i_2}\Rightarrow \qquad \omega_{i_1i_29}
&=&-{1 \over 144} \delta_{i_1i_2} \phi^{j_1j_2j_3j_4}
F^\rep{1}_{j_1j_2j_3j_4} +{1 \over 12} \phi_{(i_1}{}^{j_1 j_2 j_3}
F^{\rep{1}+\rep{35}}_{i_2) j_1 j_2 j_3} +\frac{1}{2}F_{+-i_1i_2}
\nn (\nabla_9 \Omega)_{+i} \Rightarrow \qquad \omega_{99i} &=& -{1
\over 18} \phi_i{}^{j_1 j_2 j_3} F^\rep{8}_{9j_1 j_2 j_3}- {1
\over 3}F_{+-9i} \ . \eea

We now turn to the conditions arising from derivatives of
$\Sigma$. {}From ({\ref{dkthree}}) we obtain \bea
\label{eqn:omsev} {1 \over 5!} \epsilon_{i_1 i_2 i_3}{}^{j_1 j_2
j_3 j_4 j_5} (d \phi)_{j_1 j_2 j_3 j_4 j_5} =-{2 \over 3} F_{+-9j}
\phi^j{}_{i_1 i_2 i_3}+{2 \over 3}F_{i_1 i_2 i_3 9}-{1 \over
2}F_{j_1 j_2 9 [i_1} \phi^{j_1 j_2}{}_{i_2 i_3]} \ . \nn \eea In
fact this equation fixes $\omega^\rep{7}_{ij_1j_2}\equiv
\frac{1}{4}(\omega_{ij_1j_2}-\frac{1}{2}\phi_{j_1j_2}{^{k_1 k_2}}
 \omega_{ik_1 k_2})$.
To see this define \be \psi_{i_1i_2i_3} \equiv {1 \over 5!}
\epsilon_{i_1i_2i_3}{}^{j_1 j_2 j_3 j_4 j_5} (d \phi)_{j_1 j_2 j_3
j_4 j_5} \ee then using \p{idtwo} we deduce \bea \psi_{i_1 i_2
i_3} &=& - \phi^k{}_{i_1 i_2 i_3} \omega^j{}_{jk}+3 \phi_{j_1j_2
[i_1 i_2} \omega^{j_1j_2}{}_{i_3]}\nn &=& - \phi^k{}_{i_1 i_2 i_3}
\omega^\rep{7}{}^j{}_{jk}+3 \phi_{j_1j_2 [i_1 i_2}
\omega^\rep{7}{}^{j_1j_2}{}_{i_3]} \ . \eea This expression can be
inverted to give\footnote{ We note in passing that this expression
gives a formula for minus the intrinsic con-torsion of a general
$Spin(7)$-structure in eight dimensions.} \bea\label{invexp}
\omega^\rep{7}_{ij_1j_2} &=&{1 \over 16} \psi_{ij_1j_2} +{1 \over
48} \psi_{k_1 k_2 k_3} \delta_{i[j_1} \phi^{k_1 k_2 k_3}{}_{j_2]}
- {1 \over 32} \psi_{ik_1 k_2} \phi^{k_1 k_2}{}_{j_1j_2}\nn &+&{1
\over 16} \psi_{k_1 k_2 [j_1} \phi^{k_1 k_2}{}_{j_2]i} \eea and
hence in terms of the gauge field strength \bea
\omega^\rep{7}_{ij_1j_2} &=& -{1 \over
72}\delta_{i[j_1}\phi_{j_2]}{}^{k_1 k_2 k_3}F_{k_1 k_2 k_39} -{1
\over 24} \phi_{j_1 j_2}{}^{k_1k_2} F_{ik_1 k_29} +{1 \over 24}
\phi_{i[j_1}{}^{k_1 k_2}F_{j_2]k_1 k_29} \nn &-&{1 \over
12}\delta_{i[j_1}F_{j_2]+-9}
+\frac{1}{24}\phi_{ij_1j_2}{}^kF_{k+-9} +{1 \over 12}F_{ij_1j_29}
\ . \eea

In later calculations it will be useful to note that the totally
anti-symmetric part of $\omega^\rep{7}_{j_1j_2j_3}$ can be written
as: \be \omega^\rep{7}_{[j_1j_2j_3]}
=-\frac{1}{16}\psi^\rep{8}_{j_1j_2j_3}
+\frac{1}{12}\psi^\rep{48}_{j_1j_2j_3} \ . \ee This expression and
\p{invexp} imply that  $\omega^\rep{7}_{j_1j_2j_3}$ is fixed by
the totally anti-symmetric part.

Finally from ({\ref{covc}}) we obtain \bea
(\nabla_{+}\Sigma)_{+i_1 i_2 i_3 i_4}\Rightarrow \qquad
F^\rep{7}_{+9i_1i_2} &=& 2 \omega^\rep{7}_{+i_1i_2} \nn
(\nabla_{-}\Sigma)_{+i_1 i_2 i_3 i_4}\Rightarrow \qquad
\omega^\rep{7}_{-ij} &=&0 \nn (\nabla_{9}\Sigma)_{+i_1 i_2 i_3
i_4}\Rightarrow \qquad \omega^\rep{7}_{9i_1i_2} &=&-{1 \over 72}
\phi^{j_1 j_2 j_3}{}_{i_1} F^\rep{7}_{i_2 j_1 j_2 j_3}-{1 \over
6}F^\rep{7}_{+-i_1i_2} \ . \eea

The conditions we have derived for the geometry and four-form to
admit null Killing spinors are in fact sufficient, as we shall
show in the next subsection. The careful reader will notice that
the components $F^{\rep{48}}_{+i_1i_2i_3}$, $F^{\rep{21}}_{+9 i_1
i_2}$ and $F^{\rep{27}}_{i_1..i_4}$ have not been constrained at
all. The reason for this is, as we shall see, that these
components of the field strength drop out of the Killing spinor
equation. Note that a similar phenomenon was observed for timelike
Killing spinors in \cite{gauntpakis}.

\subsection{Sufficiency}
We would like to show that the conditions derived in the last
sub-section are sufficient for the existence of null Killing
spinors, satisfying the projections \p{projs}. Let us first derive
some useful identities. Using the fact that \be {1 \over 4!}
\phi_{i_1 i_2 i_3 i_4} \Gamma^{i_1 i_2 i_3 i_4} \epsilon = 14
\epsilon \ee we obtain
\begin{eqnarray}
\Gamma_i\e&=&\frac{1}{42}\phi_i{}^{j_1j_2j_3}\Gamma_{j_1j_2j_3}\e\nn
\Gamma_{i_1i_2}\e&=&-\frac{1}{6}\phi_{i_1i_2}{}^{j_1j_2}\Gamma_{j_1j_2}
\e\nn \Gamma_{i_1i_2i_3}\e
&=&-\frac{1}{4}\phi_{[i_1i_2}{}^{j_1j_2}\Gamma_{i_3]j_1j_2}\e
=-\phi_{i_1 i_2 i_3}{^j}\Gamma_j\e\nn
\Gamma_{i_1i_2i_3i_4}\e&=&\frac{1}{4!}\e_{i_1i_2i_3i_4}{}^{j_1j_2j_3j_4}
\Gamma_{j_1j_2j_3j_4}\e=\phi_{[i_1i_2i_3}{}^j\Gamma_{i_4]j}\e
+\phi_{i_1i_2i_3i_4}\e\nn
\Gamma_{i_1i_2i_3i_4i_5}\e&=&5\phi_{[i_1i_2i_3i_4}\Gamma_{i_5]}\e
\ .
\end{eqnarray}
In particular we see that
\begin{eqnarray}\label{spinconds}
\Gamma^{\rep{21}}_{i_1 i_2}\e&=&0\nn \Gamma^{\rep{48}}_{i_1 i_2
i_3}\e&=&0\nn \Gamma^{\rep{27}}_{i_1 i_2 i_3 i_4}\e&=&0\nn
\Gamma^{\rep{35}}_{i_1 i_2 i_3 i_4}\e&=&0 \ .
\end{eqnarray}

Now the Killing spinor equation is
$\nabla_{\a}\e+\frac{1}{288}M_{\a}\e=0$ where \be M_\alpha \equiv
\Gamma_\alpha{}^{\nu_1 \nu_2 \nu_3 \nu_4} F_{\nu_1 \nu_2 \nu_3
\nu_4}-8 \Gamma^{\nu_1 \nu_2 \nu_3} F_{\alpha \nu_1 \nu_2 \nu_3} \
. \ee Using the above identities, together with the constraints
$\Gamma^9 \epsilon = \epsilon$ and $\Gamma^+ \epsilon =0$, and the
expressions for $F$ presented in the last subsection, it is
straightforward to show that \bea \label{eqn:mexps} M_- \epsilon
&=& 0 \nn M_+ \epsilon &=& \big[ \Gamma^- \big( \phi^{i_1 i_2 i_3
i_4}F_{i_1 i_2 i_3 i_4}+4 \Gamma^q \phi^{i_1 i_2 i_3}{}_q F_{9i_1
i_2 i_3}+48 \Gamma^q F_{+-9q} \big) \nn
 &-&36 \Gamma^{i_1 i_2}F_{+9i_1 i_2}+12 \Gamma^q \phi^{i_1 i_2 i_3}{}_q
F_{+i_1 i_2 i_3} \big] \epsilon \nn M_9 \epsilon &=& \big[
\Gamma^{ij} \big(F_{i_1 i_2 i_3 [i}\phi^{i_1 i_2 i_3}{}_{j]}+12
F_{+-ij} \big) +8 \Gamma^j \big(\phi^{i_1 i_2 i_3}{}_j F_{9i_1 i_2
i_3}-6F_{+-9j}\big) \nn &+&\phi^{i_1 i_2 i_3 i_4}F_{i_1 i_2 i_3
i_4} \big] \epsilon \nn M_i \epsilon &=& \big[ 72\Gamma^- \Gamma^j
F_{-9ij} -4 \phi_i{}^{j_1 j_2 j_3}F_{9j_1 j_2 j_3}-48F_{+-i9} \nn
&+& \Gamma^j \big( \delta_{ij} \phi^{j_1 j_2 j_3 j_4}F_{j_1 j_2
j_3 j_4}-4 \phi_i{}^{k_1 k_2 k_3}F_{jk_1 k_2 k_3} -12
\phi_{ij}{}^{k_1 k_2}F_{+-k_1 k_2} \nn &+&8\phi^{k_1 k_2 k_3}{}_j
F_{i k_1 k_2 k_3}-48F_{+-ij} \big) + \Gamma^{j_1 j_2}
\big(3F_{9k_1 k_2 [j_1}\phi_{j_2] i}{}^{k_1 k_2}-\delta_{i
[j_1}\phi_{j_2] k_1 k_2 k_3} F_9{}^{k_1 k_2 k_3} \nn &+&24
\delta_{i [j_1}F_{j_2]+-9}-24 F_{i9j_1 j_2} \big) \big] \epsilon \
. \eea

{}From these expressions and using \p{spinconds} it is clear that
$F^{\rep{48}}_{+i_1i_2i_3}$, $F^{\rep{21}}_{+9 i_1 i_2}$ and
$F^{\rep{27}}_{i_1..i_4}$ do not appear in $M_\mu \epsilon$ and
hence these components are not fixed by the Killing spinor
equation. Moreover, by making use of these expressions, and using
$V_{ij}\Gamma^{ij}\e=V^\rep{7}_{ij}\Gamma^{ij}\e$, we find that
\be {1 \over 4} \omega_{\mu \alpha \beta} \Gamma^{\alpha \beta}
\epsilon +{1 \over 288} M_\mu \epsilon =0 \ee and hence the
Killing spinor equation simplifies to \be d \epsilon =0 \ . \ee
Hence the Killing spinor is constant and constrained by \p{projs}.

\subsection{Summary}

We have derived the necessary and sufficient conditions on
configurations admitting null Killing spinors. Here we shall
summarise the results. Conceptually, it is clearest to separate
the conditions into a set of restrictions on the spin connection,
which are restrictions on the intrinsic torsion of the
$(Spin(7)\ltimes\bR^8)\times \bR$ structure, and a set of
conditions that determine the field strength in terms of the
geometry. In the frame \be ds^2=2e^+e^-+e^ie^i+e^9e^9 \ee where
$i=1,\dots ,8$ and
$\phi=\frac{1}{4!}\phi_{i_1i_2i_3i_4}e^{i_1i_2i_3i_4}$ is the
Spin(7) invariant four-form given in \p{phidef}, we have found the
following constraints on the spin connection
\begin{eqnarray}\label{spinrestriction}
\omega_{(\a\b)-}&=&0\nn \omega^{\bf 7}_{[ij]-}&=&0\nn
\omega_{i9-}&=&0\nn \omega_{-9i}&=&0\nn \omega^{\bf
7}_{-ij}&=&0\nn \omega_{+-9}&=&-\frac{1}{4}\omega^i{_{i9}}\nn
\omega^{\bf 7}_{9ij}&=&-\omega^{\bf 7}_{[ij]9}\nn
\omega_{99i}-6\omega_{i-+}&=&
-\frac{4}{3}\phi_{i}{^{j_1j_2j_3}}\omega^\rep{7}_{j_1j_2j_3} \ .
\end{eqnarray}
The bold-faced superscripts refer to $Spin(7)$ representations.
Note that the right hand side of the last term can also be written
as $-\phi_{i}{^{j_1j_2j_3}}\omega_{j_1j_2j_3} - 2\omega^j{}_{ji}$.

Given a geometry satisfying the above restrictions the field
strength is determined by,
\begin{eqnarray}\label{fstrengthrestriction}
F_{+-9i}&=&2\omega_{i-+}-\omega_{99i}\nn
F_{+-ij}&=&2\omega_{[ij]9}\nn F^{\bf 7}_{+9ij}&=&2\omega^{\bf
7}_{+ij}\nn F^{\bf
8}_{+i_1i_2i_3}&=&\frac{2}{7}\phi_{i_1i_2i_3}{^{j}}\omega_{+9j}\nn
F^{\bf 7}_{-9 ij}&=&0\nn F^{\bf 21}_{-9ij}&=&2\omega^{\bf 21}_{ij
-}\nn F_{- i_1i_2i_3}&=&0\nn F^{\bf
8}_{9i_1i_2i_3}&=&\frac{2}{7}\phi_{i_1i_2i_3}{^j}(\omega_{99j}
+\omega_{j-+})\nn F^{\bf 48}_{9i_1i_2i_3}&=&-12(\omega^{\bf
7}_{[i_1i_2i_3]})^\rep{48}\nn F^{\bf
1}_{i_1i_2i_3i_4}&=&\frac{3}{7}(\omega_{+9-})\phi_{i_1i_2i_3i_4}\nn
F^{\bf 7}_{i_1i_2i_3i_4}&=&2\phi_{[i_1i_2i_3}{^j}\omega^{\bf
7}_{i_4]j~9}\nn F^{\bf
35}_{i_1i_2i_3i_4}&=&2\phi_{[i_1i_2i_3}{^j}\omega^\rep{35}_{i_4]
j9} \ ,
\end{eqnarray}
where
$\omega^\rep{35}_{ij9}=\omega_{(ij)9}-\frac{1}{8}\delta_{ij}\omega^{k}
{_{k9}}$ and  $(\omega^{\bf 7}_{[ijk]})^\rep{48}$ is the
$\rep{48}$ piece of the totally anti-symmetric part of
$\omega^\rep{7}_{ijk}$. Note also that $\omega^\rep{7}_{ij9}$
denotes the $\rep{7}$ piece of $\omega_{[ij]9}$. The remaining
components of the field strength, $F^{\rep{48}}_{+i_1i_2i_3}$,
$F^{\rep{21}}_{+9 i_1 i_2}$ and $F^{\rep{27}}_{i_1..i_4}$, are
undetermined by the Killing spinor equation as shown in the
previous section, but are fixed by the Bianchi identity and gauge
field equations, which we now discuss.

\subsection{Conditions for supersymmetric solutions}

It was shown in \cite{gauntpakis} that in order for a
configuration $(g,F)$ with a null Killing spinor to also solve the
equations of motion of D=11 supergravity, it is sufficient to just
impose both the equation of motion and the Bianchi identity for
$F$ and in addition the $++$ component of Einstein's equations.
Clearly these conditions will constrain the components of $F$ not
constrained by the Killing spinor equation alone.

\section{Introducing co-ordinates}

To introduce co-ordinates, note that locally, we can choose
co-ordinates $v,u,z$ so that the vector fields dual to our chosen
frame are given by \bea \label{eqn:threevect} e^+ &=& {\partial
\over \partial v} \nn e^- &=& \alpha_1  {\partial \over \partial
v}+ \alpha_2 {\partial \over \partial u} \nn e^9 &=& \beta_1
{\partial \over \partial v} + \beta_2 {\partial \over \partial
u}+\beta_3  {\partial \over \partial z} \eea with $\alpha_2 \neq
0$, $\beta_3 \neq 0$. If the remaining co-ordinates are $x^M$,
$M=1, \dots , 8$, then as a consequence of $i_{e^+} e^i = i_{e^-}
e^i = i_{e^9} e^i=0$ we obtain \be e^i = e^i{}_M dx^M \ . \ee
Inverting ({\ref{eqn:threevect}}) we find that \bea
\label{eqn:threeinv} e^+ &=& {1 \over \alpha_2} du -{\beta_2 \over
\alpha_2 \beta_3}dz + \lambda \nn e^- &=&-{\alpha_1 \over
\alpha_2} du + dv +\big({\alpha_1 \beta_2 \over \alpha_2
\beta_3}-{\beta_1 \over \beta_3}\big) dz + \nu \nn e^9 &=&{1 \over
\beta_3} dz + \sigma \eea where $\lambda= \lambda_M dx^M$, $\nu =
\nu_M dx^M$, $\sigma = \sigma_M dx^M$. By examining the $du dv$
and $du^2$ components of the metric, it is clear that as $K$ is
Killing, $\alpha_1$ and $\alpha_2$ do not depend on $v$.
Furthermore, on examination of the $dv dz$, $du dz$ and $dz^2$
components of the metric, we also find that $\beta_1$, $\beta_2$
and $\beta_3$ must also be independent of $v$; and from the $dv
dx^M$, $du dx^M$ and $dz dx^M$ components, it is clear that $\CL_K
\lambda = \CL_K \nu = \CL_K \sigma =0$. The $dx^M dx^N$ components
of the metric then imply that $ \CL_K (e^i e^i)=0$; we shall find
it convenient to refer to the 2-parameter family of 8-manifolds
equipped with metric \be ds_8{}^2 = e^i e^i \ee as the base space
$B$. Next note that from the differential constraints
 \p{spinrestriction}
we obtain \be \CL_K e^i = \rho^i e^+ + \chi_{ij}e^j \ee where
$\chi_{ij}=-\omega_{ij-}-\omega_{-ij}$ and $\rho^i =
\omega_{+i-}-\omega_{-i+}$. Note in particular, that
$\chi_{(ij)}=0$ and $\chi^{\rep{7}}=0$. However, we also have \be
0 = \CL_K (e^i \otimes e^i) = \rho \otimes e^+ + e^+ \otimes \rho
\ee where $\rho \equiv \rho^i e^i$. Hence we must have $\rho^i=0$,
and \be \CL_K e^i = \chi_{ij} e^j \ . \ee Note that we can choose
a basis of $B$, $e'^i$ where $\CL_K e'^i=0$. To see this consider
the metric on $B$, which we denote by $h$ where \be h_{MN} \equiv
\delta_{ij} e^i{}_M(v,u,z,x) e^j{}_N(v,u,z,x) \ . \ee By the above
reasoning, $h_{MN}$ does not depend on $v$, and so on evaluating
$e^i{}_M(v,u,z,x)$ at $v=0$ we find that \be h_{MN} \equiv
\delta_{ij} (e')^i{}_M(u,z,x) (e')^j{}_N(u,z,x) \ee where \be
(e')^i{}_M(u,z,x) \equiv  e^i{}_M(0,u,z,x) \ . \ee It is clear
that $e'$ defines a basis of $B$ for which $\CL_K e'^i=0$. In
fact, the coefficients of $\phi$ are also constant in this basis.
This is because $\CL_K \phi=0$ as a consequence of
$\chi^\rep{7}=0$. Hence the components $\phi_{M_1 M_2 M_3 M_4}$ do
not depend on $v$. However, we also have \be \phi_{M_1 M_2 M_3
M_4} = \phi_{i_1 i_2 i_3 i_4} e^{i_1}{}_{M_1} (v,u,z,x)
 e^{i_2}{}_{M_2} (v,u,z,x)
 e^{i_3}{}_{M_3} (v,u,z,x)  e^{i_4}{}_{M_4} (v,u,z,x) \ .
\ee So, by the same reasoning as used above, on evaluating
$e^i{}_M(v,u,z,x)$ at $v=0$, we must have \be \phi_{M_1 M_2 M_3
M_4} = \phi_{i_1 i_2 i_3 i_4} e'^{i_1}{}_{M_1} (u,z,x)
 e'^{i_2}{}_{M_2} (u,z,x)
 e'^{i_3}{}_{M_3} (u,z,x)  e'^{i_4}{}_{M_4} (u,z,x)
\ee which implies that the components of $\phi$ in the basis
$e'^i$ are identical to those in the basis $e^i$. Hence, without
loss of generality we can drop the primes and work with a basis
$e^i$ for which {\it{both}} $\CL_K e^i=0$ {\it{and}} the
components of $\phi$ are of the canonical form given in
({\ref{phidef}}).

To continue we will introduce a more convenient notation: \bea
\label{newframe} e^+ &=& L^{-1} (du + A dz + \lambda) \nn e^-
&=&dv +\frac{1}{2}{\cal F}du + B dz + \nu \nn e^9 &=& C(dz +
\sigma) \nn e^i &=& e^i_Mdx^M \eea where the Lie-derivative of the
functions $L, {\cal F}, A, B, C$ and the one-forms $\lambda,\nu,
\sigma, e^i$ with respect to $K$ all vanish i.e. they are all
functions of $u,x^M$ and $z$ only.

It is convenient to define some notation. For a  q-form on the
base manifold \be \Theta = {1 \over q!} \Theta_{M_1 \dots M_q}
dx^{M_1} \wedge \dots \wedge dx^{M_q} \ee satisfying ${\cal L}_K
\Theta=0$, we define the restricted exterior derivative \be
\label{eqn:notata} {\tilde d} \Theta \equiv {1 \over (q+1)!} (q+1)
{\partial \over \partial x^{M_1}} \Theta_{M_2 \dots M_{q+1}}
dx^{M_1} \wedge \dots \wedge dx^{M_{q+1}} \ee and denote the Lie
derivative on such forms with respect to ${\partial \over \partial
z}$ and ${\partial \over \partial u}$ by $\partial_z$ and
$\partial_u$ respectively. We define \be \label{eqn:notatb} \CD
\Theta \equiv {\tilde d} \Theta +(A \sigma -\lambda) \wedge
\partial_u \Theta - \sigma \wedge \partial_z \Theta
\ee so that \be d \Theta = \CD \Theta +L e^+ \wedge \partial_u
\Theta +C^{-1} e^9 \wedge (\partial_z \Theta -A \partial_u \Theta)
\ . \ee We also define
\begin{eqnarray}
\label{eqn:mldef} M_{ij}&=& \delta_{ik}(\partial_u e^k)_{j} \nn
\Lambda_{ij}&=& \delta_{ik}(\partial_z e^k)_j \ .
\end{eqnarray}
In general,  $\Lambda$ and $M$ have no symmetry properties. Using
this notation, it is straightforward to compute the spin
connection. All of the components of the spin connection are
presented in Appendix D.

In order to examine the restrictions on the eleven dimensional
geometry imposed by the constraints in ({\ref{spinrestriction}}),
first observe that the basis ({\ref{newframe}}) contains a great
deal of gauge freedom. In general, there is not a single gauge
choice that simplifies all solutions, so it is convenient to allow
some gauge freedom in the final form of the geometry.
Nevertheless, to simplify the resulting formulae, we will work in
a gauge with  $A=0$, which can be achieved by making a shift of
the form $u \rightarrow u+f(u,z,x^M)$. Working in this gauge,
({\ref{spinrestriction}}) implies that
\begin{eqnarray}
\label{eqn:constraintva} [\CD \lambda]_{ij}^{\bf 7}&=&0 \nn
\partial_z \lambda &=&0 \nn
\big[\CD \log(C L^{-3})-\partial_z \sigma-3 \partial_u \lambda
\big]_i &=&-\frac{1}{2\cdot 4!}[\CD \phi]_{i
j_1..j_4}\phi^{j_1..j_4}\nn
\partial_z \log L&=&\frac{1}{2}\Lambda^i{_{i}}\nn
\Lambda_{[ij]}^{\bf 7}&=&0 \ .
\end{eqnarray}

Note that the last two equations can be expressed in terms of the
$Spin(7)$ structure $\phi$ as,
\begin{eqnarray}
\label{eqn:confasd}
\partial_z \phi=(\partial_{z} \log L) \phi+\Upsilon^{\bf 35}
\end{eqnarray}
where, denoting by $\Lambda^{\bf 35}$ the traceless symmetric
part, we defined
\begin{eqnarray}
\Upsilon^{\bf 35}_{i_1...i_4}\equiv -4\phi_{i[j_1 j_2
j_3}\Lambda^{\bf 35}{^i}{_{j_4]}} \ .
\end{eqnarray}
Using the terminology of {\cite{bryant}}, we recall that a
$Spin(7)$ structure satisfying ({\ref{eqn:confasd}}) is called
conformally anti-self-dual. Note that on making a conformal
re-scaling of the base metric $ds_8^2=L^{1/2}\hat{e}^i\hat{e}^i$,
({\ref{eqn:confasd}}) becomes
\begin{equation}
\partial_z \hat{\phi}= {\hat{\Upsilon}}^{\bf 35}
\end{equation}
where $\hat{\phi} =L^{-1} \phi$ and ${\hat{\Upsilon}}^{\bf
35}=L^{-1} \Upsilon^{\bf 35}$. Thus it makes sense to write the
conditions in terms of these conformally rescaled variables. We do
this in the following summary, where we also write out the
four-form field strength.

\subsection{Summary}
We have shown that coordinates $(u,v,z,x^M)$ can be chosen so that
the metric takes the form \be ds^2=2e^+e^-+ e^i e^i +e^9e^9 \ee
where \bea\label{basissum} e^+ &=& L^{-1} (du + \lambda) \nn e^-
&=&dv +\frac{1}{2}{\cal F}du + B dz + \nu \nn e^9 &=& C(dz +
\sigma) \nn e^i &=& L^{1/4} {\hat{e}}^i_Mdx^M \ . \eea The
eight-dimensional base manifold with metric $\hat e^i\hat e^i$ has
Cayley four-form $\hat\phi$ given by ({\ref{phidef}}), (with
$\phi$ replaced by ${\hat{\phi}}$ and $e^i$ replaced by
${\hat{e}}^i$). In general all quantities can depend on the
co-ordinates $(u,z,x^M)$.

Supersymmetry implies that the following constraints must hold
\bea \label{eqn:constrvb} (\CD \lambda)^{\rep{7}} &=&0 \nn
\partial_z \lambda &=& 0
\nn \big[\CD \log(C L^{{1 \over 2}})-\partial_z \sigma-3
\partial_u \lambda \big]_i&=&-{1 \over 48} [\CD {\hat{\phi}}]_{i
j_1..j_4}{\hat{\phi}}^{j_1..j_4} \nn
\partial_z \hat{\phi}&=& {\hat{\Upsilon}}^{\bf 35}
\eea where all indices are evaluated with respect to the
${\hat{e}}^i$ basis and the boldface numbers denote $Spin(7)$
irreps of forms taken with respect to the $Spin(7)$ structure
${\hat{\phi}}$. The derivative ${\cal D}$ is defined in
\p{eqn:notatb} (with $A=0$).

In addition, it is straightforward to show that the 4-form $F$ is
given by \bea \label{eqn:fexpression} F &=&e^+ \wedge e^- \wedge
e^9 \wedge (L^{-1} \CD L -C^{-1} \CD C +
\partial_u \lambda + \partial_z \sigma)
+C e^+ \wedge e^- \wedge \CD \sigma \nn &+& e^+ \wedge e^9 \wedge
\big( -L^{3/2} \hat M_{[ij]} {\hat{e}}^i \wedge {\hat{e}}^j - \CD
\nu + \CD B \wedge \sigma +{1 \over 2} \CD \cF \wedge \lambda
\big)^{\rep{7}} +L^{-1} e^- \wedge e^9 \wedge \CD \lambda \nn &-&
{1 \over 42}L^{1 \over 2} C^{-1} {\hat{\phi}}_{i_i i_2 i_3}{}^j
\big[ (\partial_u B - {1 \over 2} \partial_z \cF) \lambda+
\partial_z \nu - {\tilde{d}} B+LC^2 \partial_u \sigma \big]_j e^+
\wedge {\hat{e}}^{i_1} \wedge {\hat{e}}^{i_2} \wedge
{\hat{e}}^{i_3} \nn &-&L^{1 \over 2} e^9 \wedge {\hat{\star}}_8
\CD {\hat \phi} +{L^{1 \over 2} \over 6} {\hat{\phi}}_{i_1 i_2
i_3}{}^j \big[\partial_u \lambda \big]_j
 e^9 \wedge {\hat{e}}^{i_1} \wedge {\hat{e}}^{i_2}
 \wedge {\hat{e}}^{i_3}
+ {1 \over 2} LC^{-1} \partial_z {\hat{\phi}} \nn &+&{3 \over 14}
C^{-1} \partial_z L {\hat{\phi}} -{1 \over 24} L^{1 \over 2}C
{\hat{\phi}}^j{}_{[i_1 i_2 i_3} \CD \sigma_{i_4] j}
{\hat{e}}^{i_1} \wedge {\hat{e}}^{i_2} \wedge {\hat{e}}^{i_3}
\wedge {\hat{e}}^{i_4} +F_{{\rm unfixed}} \ , \eea where, again,
all indices are evaluated in the ${\hat{e}}^i$ basis,
${\hat{\star}}_8$ denotes the Hodge dual with respect to the
metric ${\hat{e}}^i {\hat{e}}^i$ and
${\hat{M}}_{ij}=\delta_{ik}(\partial_u \hat{e}^k)_j$. $F_{{\rm
unfixed}}$ contains the components $F^{\rep{48}}_{+i_1i_2i_3}$,
$F^{\rep{21}}_{+9 i_1 i_2}$ and $F^{\rep{27}}_{i_1..i_4}$, which
are undetermined by the Killing spinor equation.

\section{Examples}

In this section we consider some special examples of
supersymmetric geometries. These provide some concrete insight
into the $(Spin(7)\ltimes\bR^8)\times \bR$-structures that we have
shown supersymmetry dictates. We first consider the resolved
membranes of \cite{Cvetic:2000mh}, presenting a new generalisation
involving the addition of a gravitational wave, followed by the
basic fivebrane solution.

\subsection{Membranes and their resolution}

The elementary membrane and fivebrane solutions of $D=11$
supergravity admit 16 Killing spinors. Some of these are timelike
and the corresponding $SU(5)$ structure was displayed in
\cite{gauntpakis}. However some of the spinors are null so these
solutions also fall in the null case that we are studying here.
Let us focus first on the membrane. The metric and field strength
for this solution are given by,
\begin{eqnarray}\label{M2brane}
ds^2&=&H^{-2/3}[-dt^2+(dx^\sharp)^2+dz^2]+H^{1/3}ds^2(\bR^8)\nn
F&=&dt\wedge dx^\sharp \wedge dz \wedge d(H^{-1})
\end{eqnarray}
where the gauge equations imply that $H$ is a harmonic function on
$\bR^8$.

One can generalize this solution by replacing the space transverse
to the membrane (which is $\bR^8$ in ({\ref{M2brane}})) by any
$Spin(7)$ holonomy manifold \cite{Duff:1995wk}. An additional
generalisation leads to the ``resolved membrane" solutions of
\cite{Cvetic:2000mh}. To see how this latter solution is related
to our construction of solutions of eleven dimensional
supergravity, we introduce a null frame
\begin{eqnarray}\label{M2nullframe}
e^+&=&\frac{H^{-2/3}}{\sqrt{2}}(-dt+dx^\sharp)\nn
e^-&=&\frac{1}{\sqrt{2}}(dt+dx^\sharp)\nn e^9&=&H^{-1/3}dz \nn
e^i&=&H^{1/6} {\hat{e}}^i
\end{eqnarray}
where $ds^2(M_8)=\hat{e}^i \hat{e}^i$ is a $Spin(7)$ holonomy
metric. Recall that this implies that $\hat \phi$ is closed,
${\tilde{d}} {\hat{\phi}}=0$. Both $H$ and $\hat{e}^i$ are
independent of $t,x^\sharp,z$. On setting $v={1 \over
\sqrt{2}}(t+x^\sharp)$, $u={1 \over \sqrt{2}} (-t+x^\sharp)$ it is
clear that ({\ref{M2nullframe}}) corresponds to the null basis
given in \p{basissum} with $L=H^{2 \over 3}$, $A={\cal{F}}=B=0$,
$\lambda=\nu=\sigma=0$, $C= H^{-{1 \over 3}}$ and $e^i = H^{1
\over 6} {\hat{e}}^i{}_M dx^M$. It is then simple to check that
the constraints required for supersymmetry \p{eqn:constrvb} are
satisfied.

The expression for the field strength given in \p{eqn:fexpression}
takes the form \be \label{Feqn} F = H^{-1} e^+ \wedge e^- \wedge
e^9 \wedge dH +\hat F^\rep{27} \ee where we have allowed for a
piece, ${\hat{F}}^\rep{27}={1 \over 4!}{\hat{F}}^\rep{27}{}_{i_1
i_2 i_3 i_4} \hat{e}^{i_1} \wedge \hat{e}^{i_2} \wedge
\hat{e}^{i_3} \wedge \hat{e}^{i_4}$, in the {\bf 27} on $M_8$ that
is not fixed by supersymmetry.

Imposing the gauge equations of motion we find that ${\hat{F}}^{\bf 27}$
must be closed (and hence harmonic) while the equation for $H$
becomes
\begin{equation}\label{resolvedH}
{\hat{\nabla}}^2 H=-\frac{1}{2}|{\hat{F}}^{\bf 27}|^2
\end{equation}
where ${\hat{\nabla}}^2$ is the laplacian on $M_8$ and the norm of
${\hat{F}}^{\bf 27}$ is taken in the metric $ds^2(M_8)$. The $++$
component of the Einstein equations imposes no further
restriction. Such ``resolved membrane" solutions were constructed
in \cite{Cvetic:2000mh} although there the issue of supersymmetry
was not discussed and the internal component of the field strength
was only constrained to be self-dual. The supersymmetry of such
solutions was discussed in \cite{Becker:2000jc,Cvetic:2000db}. The
condition on the internal flux given in these papers is exactly
the statement that it should belong to the {\bf 27} of $Spin(7)$.

A generalization of the membrane solution preserving just 1/32
supersymmetry was constructed in \cite{Gomis:2003zw}. The
generalization involved replacing $e^- \to e^- +(1/2){\cal
F}du+\nu$ in (\ref{M2nullframe}) where ${\cal F}, \nu$ depend just
on the coordinates on the $Spin(7)$ manifold. This is a
supersymmetric solution provided that ${\cal F}$ is harmonic and
\be\label{lilt} d*_8d\nu=0 \ee on $M_8$. This was interpreted as
adding a wave along the membrane although the wave ``profile"
${\cal F}$ was smeared in the direction $u$.

The solutions of  \cite{Cvetic:2000mh,Gomis:2003zw} can
be combined to yield a
new, more general solution, by including both the term
${\hat{F}}^{\bf 27}$ and ${\cal F},\nu$
where now we allow ${\cal F}={\cal F}(u,z,x^M)$ and maintain
$\nu=\nu(x^M)$. In addition to ${\hat{F}}^{\bf 27}$,
we let $F_{\rm unfixed}$ in
\p{eqn:fexpression} also contain the piece: $F^{\rep {21}}_{+9ij}=
-(d\nu)^{\rep{21}}_{ij}$.
The gauge equations again
imply that $\nu$ satisfies \p{lilt}
and ${\hat{F}}^{\bf 27}$ is harmonic on $M_8$ and $H$
satisfies (\ref{resolvedH}),
while the $++$ component of the Einstein equations gives,
\be \label{eqn:curlyfeqn} {\hat{\nabla}}^2 {\cal{F}}
+ H {\partial^2 {\cal{F}} \over \partial z^2} =0 \ .
\ee

In general these solutions will preserve 1/32 supersymmetry. Notice
that the dependence
of ${\cal F}$ on $u$ is not fixed. This is as expected since a
supersymmetric wave is allowed to have
an arbitrary profile.
Note also that there is a special case when the $Spin(7)$
manifold is a product of two hyper-K\"ahler manifolds and $\nu=0$;
for this case the
resulting solutions are special cases of a class of solutions
presented in \cite{Gauntlett:1998kc}.

\subsection{The fivebrane}

The metric for the basic fivebrane solution can be written as \be
\label{eqn:mvmet} ds^2 = H^{-{1 \over
3}}[-dt^2+(dx^\sharp)^2]+H^{2 \over 3} dz^2+ds_8^2 \ee where \be
\label{eqn:eightman} ds_8^2 = H^{-{1 \over 3}} \big[ (dx^1)^2+
\dots + (dx^4)^2 \big] + H^{{2 \over 3}} \big[ (dx^5)^2+ \dots +
(dx^8)^2 \big] \ee and $H=H(z,x^5,x^6,x^7,x^8)$. On setting  $v={1
\over \sqrt{2}} (t+x^\sharp)$, $u={1 \over \sqrt{2}}(-t+x^\sharp)$
it is clear that ({\ref{eqn:mvmet}}) corresponds to the null basis
given in \p{basissum} with $C=L=H^{1 \over 3}$, $A={\cal{F}}=B=0$,
$\lambda=\nu=\sigma=0$, and we split the base indices on the
8-manifold via $e^i = \{ e^a , e^p \}$ for $a,b =1, \dots, 4$ and
$p,q = 5, \dots, 8$. The vielbein on the 8-dimensional base
${\hat{e}}^i$ is therefore \bea \label{eqn:eightsplit} {\hat{e}}^a
&=& H^{-1 /4} dx^a \nn {\hat{e}}^p &=& H^{{1/4}} dx^p \ . \eea
$\hat\phi$ is given by \bea \label{phidefb}
   -{\hat{\phi}} &=& H^{-1} dx^{1234} +H dx^{5678} + \big(
dx^{1256}+dx^{1278}+dx^{3456}+dx^{3478}
      + dx^{1357}
\nn &-&dx^{1368}-dx^{1458}-dx^{1467}-dx^{2358}
          -dx^{2367}-dx^{2457}+dx^{2468} \big)
\eea

It is then straightforward to show that the constraints given in
 ({\ref{eqn:constrvb}}) are satisfied.
In addition it is straightforward to show that \p{eqn:fexpression}
gives \be F=- dz \wedge \star_4 (\nabla_p H
dx^p)+\frac{1}{14}H^{-1}
\partial_zH\hat \phi
+\frac{H^{-1}}{2}\partial_z H \hat e^{1234}
-\frac{H^{-1}}{2}\partial_z H \hat e^{5678} +F_{unfixed} \ee where
$\nabla_p = {\partial \over \partial x^p}$ and $\star_4$ denotes
the Hodge dual on $\bR^4$ equipped with metric \be
ds_4^2=(dx^5)^2+ (dx^6)^2+(dx^7)^2 + (dx^8)^2 \ee and positive
orientation fixed with respect to $dx^5 \wedge dx^6 \wedge dx^7
\wedge dx^8$. However, unlike the case of the simple
(non-resolved) M2-brane, in order to recover the standard
expression for the components $F_{i_1 i_2 i_3 i_4}$, it is
necessary to include a contribution from $F^{\rep{27}}{}_{i_1 i_2
i_3 i_4}$ which is not fixed by the supersymmetry. This term is
given by \be F^\rep{27} =-{1 \over 14} H^{-1} \partial_z H
\big({\hat{\phi}} +7 {\hat{e}}^1 \wedge {\hat{e}}^2 \wedge
{\hat{e}}^3 \wedge {\hat{e}}^4+7 {\hat{e}}^5 \wedge {\hat{e}}^6
\wedge {\hat{e}}^7 \wedge {\hat{e}}^8 \big) \ . \ee Hence the
field strength is given by \be F = - \star_5 dH \ee where
$\star_5$ denotes the Hodge dual on $\bR^5$ equipped with metric
\be ds_5^2= dz^2+ (dx^5)^2+ (dx^6)^2+(dx^7)^2 + (dx^8)^2 \ee and
positive orientation fixed with respect to $dz \wedge dx^5 \wedge
dx^6 \wedge dx^7 \wedge dx^8$. The Bianchi identity implies that
$H$ is harmonic on $\bR^5$. The field equations for the four-form
and the $++$ component of the Einstein equations lead to no
further conditions, and we see that we have recovered the
fivebrane solution.

\section{Configurations with Vanishing Flux}
When the flux vanishes, the local form of the most general
supersymmetric configuration was written down in \cite{bryant}. It
is interesting to recover this result from our more general
results. One approach is to use the co-ordinates and frame
introduced in section 4, set $F=0$ in \p{eqn:fexpression} and then
analyse the resulting metric. However, we find it easier to obtain
the result of \cite{bryant} by introducing co-ordinates afresh, as
we now explain.

If the flux vanishes, from \p{dk} we note that $K$, $\Omega$ and
$\Sigma$ are closed. In particular, there exists (at least
locally)  functions $u$ and $v$, such that as a 1-form \be e^+ =du
\ee and as a vector \be e^+ = {\partial \over \partial v} \ . \ee
Furthermore, as $e^+ \wedge de^9=0$ as a consequence of $d
\Omega=0$, we note that there must exist functions $z$ and $P$
such that \be e^9 = dz+P du \ . \ee Next consider the $e^i$, and
$e^-$. In these co-ordinates, in general we have \be e^i = e^i_M
dx^M + X^i du + Y^i dz \ee and \be e^- = dv + p_1 du +p_2 dz +
{\nu_2} \ . \ee By making a (generally $u$, $z$ dependent)
co-ordinate transformation of the $x^M$ we can work in
co-ordinates for which $Y^i=0$. Next, observe that by making a
basis rotation of the form given in ({\ref{eqn:basisrot}}) we can
work in a basis for which $X^i=0$ and $P=0$. Hence, we have shown
that if the flux vanishes, we can without loss of generality take
the basis ({\ref{newframe}}) with $L=C=1$, $A=0$,
$\lambda=\sigma=0$. By making a shift in $v$ we can also set
$B=0$. Observe that closure of $\Sigma$ implies that $\partial_z
\phi=0$ and hence in particular, $\partial_z (e^i e^i)=0$. By the
same reasoning which was used in section four to demonstrate that
$e^i$ could be chosen to be independent of $v$, we can, without
loss of generality, choose a basis $e^i$ for which $\partial_z
e^i=0$ (together with ${\cal L}_K e^i=0$). Note also that ${\tilde
d} \phi=0$.

In fact we can also set $\nu=0$. To see this note that from the
vanishing of $\omega_{+9i}$ we must have $\partial_z \nu=0$. Hence
by making a basis transformation of the form given in
({\ref{eqn:basisrot}}) (with $\alpha=0$, $Q=1$ and $p_i = \nu_i$),
we can remove the $\nu$ term from $e^-$ at the expense of adding a
$\nu^i du$ term to $e^i$. However, as $\nu^i$ has no
$z$-dependence, we can remove this term by making a
$z$-independent co-ordinate transformation of the $x^M$.

To summarize, when the flux vanishes, we can without loss of
generality work in a null basis with \bea e^+ &=& du \nn e^- &=&
dv +{1 \over 2} {\cal F} du \nn e^9 &=& dz \nn e^i &=& e^i{}_M
dx^M \eea with ${\cal F}={\cal F}(u,x,z)$ and
$e^i{}_M=e^i{}_M(u,x)$. In this basis, ${\tilde d \phi}=0$, so
$\phi$ is covariantly constant with respect the the Levi-Civita
connection on the base manifold. Although $\phi$ does not have any
$z$-dependence, it does generically have a dependence on $u$. In
particular, using ${\omega^\rep{7}}_{+ij}=0$ it follows that
$M^{\rep{7}}{}_{[ij]}=0$ and hence \be
\partial_u \phi = {\cal{T}} \phi + \Psi^{\rep{35}}
\ee where ${\cal{T}}=(1/2)M_{ii}$ i.e. $\phi$ is conformally
anti-self-dual\footnote{ It is interesting to compare this with a
generic solution, with $F \neq 0$, where $\phi$ is not conformally
anti-self dual with respect to $\partial_u$; rather, $\phi$ is
conformally anti-self-dual with respect to $\partial_z$.}. Given
this metric, ${\cal{F}}$ is fixed by the $++$ component of the
Einstein equations. Hence we have recovered the result of
\cite{bryant}.

\section{Conclusions}

In this paper we have completed the classification initiated in
\cite{gauntpakis} of solutions of eleven dimensional supergravity
preserving 1/32 of the supersymmetry. Just as in the case of
simpler, lower-dimensional supergravities, this classification
provides an interesting and promising tool for the generation of
new solutions. In addition, we have shown how several previously
known solutions, such as the zero-flux solution of \cite{bryant},
the fivebrane and the resolved membrane \cite{Cvetic:2000mh}
can we written in our formalism. We generalised the solutions of
\cite{Cvetic:2000mh,Gomis:2003zw} by adding a gravitational wave to
the resolved membrane and the resulting configuration preserves
just 1/32 supersymmetry.


Supersymmetric solutions of most physical interest preserve more
than one supersymmetry. Although such solutions are included in
our classification, it is clear that the presence of more linearly
independent Killing spinors imposes additional constraints on the
geometry. Note that, using different techniques, the
classification of maximally supersymmetric configurations,
preserving all 32 supersymmetries, was carried out in
\cite{figpap}. It would therefore be interesting to generalize our
construction  to accommodate additional linearly independent
Killing spinors. It might be possible, for example, to classify all
geometries preserving exactly four supersymmetries, or perhaps all
those preserving more than 1/2 of the supersymmetry.

\acknowledgments{
We would like to thank Daniel Waldram for helpful discussions.
J.~B.~G.~ was supported by EPSRC.}

\appendix
\makeatletter
\renewcommand{\theequation}{A.\arabic{equation}}
\@addtoreset{equation}{section} \makeatother
\section{Conventions}

We use the signature $(-,+,...,+)$. D=11 co-ordinate indices will
be denoted $\mu,\nu,\dots$ while tangent space indices will be
denoted by $\alpha,\beta,\dots$. The D=11 spinors we will use are
Majorana. The gamma matrices satisfy \be
\{\Gamma_\alpha,\Gamma_\beta\}=2\eta_{\alpha\beta} \ee and can be
taken to be real in the Majorana representation. They satisfy, in
our conventions,
$\Gamma_{0123456789\sharp}=\e_{0123456789\sharp}=1$. For any $M,N$
$\in$ $R(32)$ we can perform a Fierz rearrangement using:
\begin{eqnarray}
M{_{a}}{^b}N{_c}{^d} &=& \frac{1}{32}\{(NM){_a}{^d}\delta{_c}{^b}
+(N\Gamma^{\a_1}M){_a}{^d}(\Gamma_{\a_1}){_c}{^b} \nn
&-&\frac{1}{2!}(N\Gamma^{\a_1\a_2}M){_a}{^d}(\Gamma_{\a_1\a_2}){_c}{^b}
-\frac{1}{3!}(N\Gamma^{\a_1\a_2\a_3}M){_a}{^d}(\Gamma_{\a_1\a_2\a_3}){_c}
{^b} \nn &+&\frac{1}{4!}(N\Gamma^{\a_1\a_2\a_3\a_4}M){_a}{^d}
(\Gamma_{\a_1\a_2\a_3\a_4} ){_c}{^b}
+\frac{1}{5!}(N\Gamma^{\a_1\a_2\a_3\a_4\a_5}M){_a}{^d}
(\Gamma_{\a_1\a_2\a_3\a_4\a_5}){_c}{^b}\} \nn
\end{eqnarray}
where $a,b,c,d=1,\dots,32$.

\par Given a Majorana spinor $\e$ its conjugate  is given by $\bar{\e}
=\e^T C$, where $C$ is the charge conjugation matrix in D=11 and
satisfies $C^T=-C$. In the Majorana representation we can choose
$C=\Gamma_0$. An important property of gamma matrices in D=11 is
that the matrix $C\Gamma_{\a_1\a_2...\a_p}$ is symmetric for
$p=1,2,5$ and antisymmetric for $p=0,3,4$ (the cases $p>5$ are
related by duality to the above).

The Hodge star of a $p$-form $\omega$ is defined by \be
*\omega_{\mu_1\dots \mu_{11-p}}=\frac{\sqrt
{-g}}{p!}\epsilon_{\mu_1\dots \mu_{11-p}}
{}^{\nu_1\dots\nu_p}\omega_{{\nu_1\dots\nu_p}} \ee and the square
of a $p$-form via \be
\omega^2=\frac{1}{p!}\omega_{\mu_1\dots\mu_p}\omega^{\mu_1\dots\mu_p}
\ee unless otherwise stated.

\makeatletter
\renewcommand{\theequation}{B.\arabic{equation}}
\@addtoreset{equation}{section} \makeatother
\section{Algebraic Relations of D=11 Spinors}

Here we analyse the algebraic structure of the differential forms
$K,\Omega, \Sigma$ defined in \p{defforms} using Fierz identities.
This provides an alternative derivation of \p{met}-\p{phidef}
which relied on some results of \cite{bryant}.

Using Fierz identities one finds:
\begin{equation}\label{omsquare}
  \begin{array}{c}
     \Sigma^2=-6\K^2 \\
    \Omega^2=-5\K^2 \
  \end{array}
\end{equation}
\begin{equation}\label{complex}
\Omega{_{\mu_1}}{^{\sigma_1}}\Omega{_{\sigma_1}}{^{\nu_1}}
=-\K_{\mu_1}\K^{\nu_1}+\delta{_{\mu_1}}{^{\nu_1}}\K^2
\end{equation}
\begin{eqnarray}\label{sigmasigma}
\frac{1}{4!}\Sigma{_{\mu_1}}{^{\sigma_1\sigma_2\sigma_3\sigma_4}}
\Sigma{_{\sigma_1\sigma_2\sigma_3\sigma_4}}{^{\nu_1}}
&=&14\K_{\mu_1}\K^{\nu_1}-4\delta{_{\mu_1}}{^{\nu_1}}K^2
\end{eqnarray}
\begin{equation}\label{ikom}
i_{\K}\Omega=0
\end{equation}
\begin{equation}\label{iks}
i_{\K}\Sigma=\frac{1}{2}\Omega\wedge\Omega
\end{equation}
\begin{equation}\label{ikstars}
\K^{\s}(*\Sigma)_{\s\nu_1\nu_2\nu_3\nu_4\nu_5}
=\Omega{_{\nu_1}}{^{\s}}\Sigma_{\s\nu_2\nu_3\nu_4\nu_5} -12
\eta_{\nu_1 [ \nu_2} K_{\nu_3} \Omega_{\nu_4 \nu_5]}
\end{equation}
\bea\label{omsig}
 K^2\Omega\wedge\Sigma=\frac{1}{2}
\K\wedge\Omega\wedge\Omega\wedge\Omega \eea \bea \label{eqn:cayl}
\Omega_{\nu_1}{}^\rho \star \Sigma_{\rho \nu_2 \nu_3 \nu_4 \nu_5
\nu_6} = -5 \Sigma_{\nu_1 [\nu_2 \nu_3 \nu_4 \nu_5} K_{\nu_6]}+5
\eta_{\nu_1 [\nu_2}(i_K \Sigma)_{\nu_3 \nu_4 \nu_5 \nu_6]} \ .
\eea
Note that equations \p{omsquare}-\p{omsig} appeared previously in
\cite{gauntpakis} and that \p{ikstars} corrects equation (2.14) of
that reference.

 These are by no
means exhaustive, though they are in fact sufficient to deduce the
algebraic structures in both timelike and null cases. In
particular, ({\ref{complex}}) implies that $K$ cannot be
spacelike. To see this, note first that in a neighbourhood in
which $\epsilon$ is non-vanishing, $K_0 = -\epsilon^T \epsilon
\neq 0$. Without loss of generality, $K= -(\epsilon^T \epsilon)
e^0 +me^{\sharp}$. As $i_K \Omega=0$ we must have \be (\epsilon^T
\epsilon) \Omega_{0  \alpha} +m \Omega_{\sharp \alpha} =0. \ee In
particular, we find $\Omega_{0 \sharp}=0$ and
$\Omega_{0P}=-(m/\e^T \e)\Omega_{\sharp P}$, where $P,Q=1, \dots ,
9$. Then upon setting $\mu_1=\nu_1=\sharp$ in ({\ref{complex}}) we
find that \be \Omega_{\sharp P} \Omega_{\sharp}{}^P = m^2-K^2=
(\epsilon^T \epsilon)^2 \ . \ee But setting $\mu_1=P$, $\nu_1=Q$ in
\p{complex} we see that \be \label{eqn:contrada} \delta_{PQ} (m^2-
(\epsilon^T \epsilon)^2) = (\epsilon^T \epsilon)^{-2} (m^2-
(\epsilon^T \epsilon)^2) \Omega_{P \sharp} \Omega_{Q \sharp} -
\Omega_{P}{}^{L} \Omega_{QL} \ .  \ee Contracting with $\delta^{PQ}$ we
obtain \be 8 (m^2- (\epsilon^T \epsilon)^2) = - \Omega_{PQ}
\Omega^{PQ} \ . \ee This implies that $m^2 \leq (\epsilon^T
\epsilon)^2$, so $K$ must be timelike or null.

The case when $K$ is timelike has been examined in detail
in \cite{gauntpakis}. Here we shall concentrate on the case when
$K$ is null. It is therefore convenient to work in a null basis
 \be ds^2=2e^+e^- + e^P e^P \ee
with $K=e^+$. To proceed, note that $i_K \Omega=0$ implies that
\be \Omega= e^+ \wedge V + {1 \over 2} \Omega_{PQ} e^P \wedge e^Q
\ee where $V=V_P e^P$. However, as $\Omega^2=0$ it is
straightforward to see that $\Omega_{PQ}=0$, so \be
\label{eqn:omident} \Omega=e^+ \wedge V \ . \ee Setting $\mu_1=+$,
$\nu_1=-$ in ({\ref{complex}}) we also find that $V^2=1$. Note
that ({\ref{eqn:omident}}) implies that $\Omega \wedge \Omega=0$
and hence from ({\ref{iks}}) we find that $i_K \Sigma=0$, hence
\be \Sigma= e^+ \wedge \phi + {1 \over 5!} \Sigma_{P_1 P_2 P_3 P_4
P_5} e^{P_1 P_2 P_3 P_4 P_5} \ee where $\phi = {1 \over 4!}
\phi_{P_1 P_2 P_3 P_4} e^{P_1 P_2 P_3 P_4}$. However $\Sigma^2=0$,
so $\Sigma_{P_1 P_2 P_3 P_4 P_5}=0$, and hence \be
\label{eqn:sgident} \Sigma = e^+ \wedge \phi \ . \ee In addition, from
({\ref{ikstars}}) we note that
$\Omega{_{\nu_1}}{^{\s}}\Sigma_{\s\nu_2\nu_3\nu_4\nu_5}=0$ as $i_K
\star \Sigma=0$ and $K \wedge \Omega=0$. Setting $\nu_1=\nu_2=+$
we find that \be i_V \phi=0 \ . \ee Hence it is convenient to make an
8+1 split $e^P = \{ e^i , e^9 \}$ for $i,j=1, \dots , 8$ with
$V=e^9$ and $\phi = {1 \over 4!} \phi_{i_1 i_2 i_3 i_4} e^{i_1 i_2
i_3 i_4}$. In addition, setting $\nu_1=\nu_2=+$ in
({\ref{eqn:cayl}}) we note that $\phi$ is a self dual 4-form on
the 8-manifold equipped with metric $\delta_{ij}e^i e^j$, where we
take $\epsilon_{+-123456789}=-1$ with positive orientation on the
8-manifold given by $\epsilon_{12345678}=1$.

To proceed, we work in a particular basis in which $K=-\epsilon^T
\epsilon (e^0+e^{\sharp})$. By examining the expressions for $K$
and $V$ using the representation of Cliff(1,10) presented below,
we see that $\epsilon^a=0$ for $a=9 , \dots , 32$, or
equivalently \be \label{eqn:proja} \Gamma_9 \epsilon = \epsilon
\ee and \be \label{eqn:projb} (\Gamma_0 - \Gamma_{\sharp})
\epsilon =0 \ . \ee

Moreover, a direct examination of the components of  $\phi$ yields
the identity \be \phi^{i_1 i_2 i_3 j} \phi_{q_1 q_2 q_3 j}= 6
\delta^{i_1 i_2 i_3}_{q_1 q_2 q_3}-9 \phi^{[i_1 i_2}{}_{[q_1 q_2}
\delta^{i_3]}_{q_3]} \ . \ee In particular, we find that \be
\label{eqn:usefulident} \phi_{i_1 i_2 i_3 j} \phi_{i_1 i_2
i_3}{}^j=1 \ee for distinct fixed $i_1$, $i_2$, $i_3$.

It appears that there are eight degrees of freedom in the spinor
$\epsilon$. In fact there is only one degree of freedom, and a
basis $\{e^i \}$ can be chosen in which $-\phi$ takes  the
canonical form of the Cayley 4-form. To see this we shall
concentrate on the components $\phi_{146i}$, $\phi_{145i}$ and
$\phi_{168i}$. Observe from ({\ref{eqn:usefulident}}) that
$\phi_{146i} \phi_{146}{}^i=1$. Hence by rotating in the 2,3,5,7,8
directions we can arrange without loss of generality for
$\phi_{1467}=1$ and
$\phi_{1462}=\phi_{1463}=\phi_{1465}=\phi_{1468}=0$. By inspecting
the expression for $\phi_{1467}$ in terms of components of the
spinor, it is apparent that $\phi_{1467}=1$ implies that
$\epsilon^2=\epsilon^5=\epsilon^7=\epsilon^8=0$. Next consider
$\phi_{145i}$; again we have $\phi_{145i} \phi_{145}{}^i=1$. In
addition, the only non-vanishing components of $\phi_{145i}$ are
$\phi_{1452}$, $\phi_{1453}$ and $\phi_{1458}$. Hence, by rotating
in the 2,3,8 directions we can set without loss of generality
$\phi_{1458}=1$ and $\phi_{1452}=\phi_{1453}=0$. Note that such a
rotation will not change the values of $\phi_{146i}$. Moreover,
$\phi_{1458}=1$ implies that $\epsilon^6=-\epsilon^3$ and
$\epsilon^4=\epsilon^1$. Lastly consider $\phi_{168i}$; once more,
from ({\ref{eqn:usefulident}}) we have $\phi_{168i}
\phi_{168}{}^i=1$. In addition, the only non-vanishing components
of $\phi_{168i}$ are $\phi_{1682}$ and $\phi_{1683}$. Hence, by
rotating in the 2,3 directions we can set without loss of
generality $\phi_{1683}=1$, $\phi_{1682}=0$. Such a rotation
leaves unaltered the values of $\phi_{146i}$ and $\phi_{145i}$,
and $\phi_{1683}=1$ implies that $\epsilon^3=-\epsilon^1$.

To summarize, in this basis, we find that \bea \label{Psidef2}
   -\phi &= e^{1234}+e^{1256}+e^{1278}+e^{3456}+e^{3478}+e^{5678} \nn
      &\qquad + e^{1357}-e^{1368}-e^{1458}-e^{1467}-e^{2358}
          -e^{2367}-e^{2457}+e^{2468} ,
\eea and the only non-vanishing components of the spinor
$\epsilon$ are $\epsilon^1= -\epsilon^3= \epsilon^4=\epsilon^6$.
This corresponds to imposing the projections \be \label{eqn:projc}
\Gamma_{1234}\e=\Gamma_{3456}\e=\Gamma_{5678}\e=\Gamma_{1357}\e=-\e \ .
\ee In summary, we see that we have  rederived equations
\p{met}-\p{phidef}.

\subsection{An explicit representation of Cliff(10,1)}

In order to compute some of the Fierz identities and algebraic
relations satisfied by the various bi-linears, it is useful to have
an explicit representation for Cliff(10,1). We recall the
representation given in \cite{Figueroa-O'Farrill:1999tx}. In
particular, let $L_i$ denote left multiplication by the imaginary
octonions on the octonions, for $i=1, \dots , 7$. Explicitly, if
$e_i$ for $i=1, \dots , 7$ denote the imaginary unit octonions,
then we take \be e_i . e_j = - \delta_{ij}+c_{ijk}e_k \ee where
$c_{ijk}$ is totally skew and has non-vanishing components fixed
(up to permutation of indices) by \be
c_{124}=c_{137}=c_{156}=c_{235}=c_{267}=c_{346}=c_{457}=1 \ . \ee

Then it is straightforward to construct the representation of
Cliff(8,0) by defining the following $16 \times 16$ real block
matrices \bea {\hat{\Gamma}}_i &=& \pmatrix{0 \quad \ \ \ \ \ \
L_i \cr L_i \quad \ \ \ \ \ \ 0} \nn {\hat{\Gamma}}_8 &=&
\pmatrix{0 \quad \ \ \ -1 \cr 1 \ \ \ \ \ \ \quad 0} \eea for
$i=1, \dots , 7$. The representation of Cliff(10,1) is then
obtained by defining the following $32 \times 32$ real block
matrices \bea \Gamma_i &=& \pmatrix{0 \quad \ \ \
-{\hat{\Gamma}}_i \cr {\hat{\Gamma}}_i \quad \ \ \ \ \ \ 0} \nn
\Gamma_9 &=& \pmatrix{1 \quad \ \ \ \ \ \ 0 \cr 0 \quad \ \ \ \
-1} \nn \Gamma_{\sharp} &=& \pmatrix{0 \quad \ \ \ \ \ \ \ 1 \cr 1
\quad \ \ \ \ \ \ \ \ 0} \eea for $i=1, \dots , 8$ and \be
\Gamma_0 =- \Gamma_{123456789\sharp} \ . \ee

\makeatletter
\renewcommand{\theequation}{C.\arabic{equation}}
\@addtoreset{equation}{section} \makeatother
\section{Spin(7) Identities}
The Spin(7) 4-form $\phi$ satisfies the following identities,
which, as far as we know, are new: \be\label{idone} 42 \phi_{[i_1
i_2}{}^{[j_1 j_2} \delta^{j_3 j_4]}_{i_3 i_4]} + \phi_{i_1 i_2 i_3
i_4}\phi^{j_1 j_2 j_3 j_4}-3 \phi_{[i_1 i_2}{}^{[j_1 j_2}
\phi^{j_3 j_4]}{}_{i_3 i_4]} +2 \phi^{[j_1}{}_{[i_1 i_2
i_3}\phi_{i_4]}{}^{j_2 j_3 j_4]}=0 \ee and \bea\label{idtwo} {1
\over 4!} \epsilon_{i_1 i_2 i_3 i_4}{}^{j_1 j_2 j_3 j_4} &=& {1
\over 168} \phi_{i_1 i_2 i_3 i_4} \phi^{j_1 j_2 j_3 j_4} +{3 \over
28} \phi_{[i_1 i_2}{}^{[j_1 j_2} \phi^{j_3 j_4]}{}_{i_3 i_4]} +{2
\over 21} \phi^{[j_1}{}_{[i_1 i_2 i_3}\phi_{i_4]}{}^{j_2 j_3 j_4]}
\nn \eea We also have the well known identities \bea
\label{eqn:spinvii}
   \phi^{i_1i_2i_3k}\phi_{j_1j_2j_3k}
      &=&6\delta^{i_1i_2i_3}_{j_1j_2j_3} -9 \phi^{[i_1i_2}{}_{[j_1j_2}
      \delta^{i_3]}_{j_3]} , \nn
   \phi^{i_1i_2k_1k_2}\phi_{j_1j_2k_1k_2}
      &=&12\delta^{i_1i_2}_{j_1j_2} -4 \phi^{i_1i_2}{}_{j_1j_2} , \nn
   \phi^{ik_1k_2k_3}\phi_{jk_1k_2k 3}
      &=&42\delta^{i}_{j} .
\eea

Given these identities one can show that \bea
\phi^k{}_{[i_1i_2i_3}\aomega^\rep{21}_{i_4]k}&=&0\nn
\phi_{[i_1i_2}{}^{j_1j_2}\aomega_{i_3]j_1j_2}&=&(-4\aomega^\rep{8}
+\frac{2}{3}\aomega^\rep{48})_{i_1i_2i_3}\nn
\phi_{[i_1i_2}{}^{j_1j_2}\aomega_{i_3i_4]j_1j_2}&=&
(-4\aomega^\rep{1}-2\aomega^\rep{7}+\frac{2}{3}
\aomega^\rep{27})_{i_1i_2i_3i_4}\nn
\phi_{[i_1i_2}{}^{j_1j_2}\phi_{i_3i_4]}{}^{j_3j_4}
\aomega_{j_1j_2j_3j_4}&=&
(28\aomega^\rep{1}-12\aomega^\rep{7}+\frac{28}{3}\aomega^\rep{27}
-4\aomega^\rep{35})_{i_1i_2i_3i_4}\nn
\phi_{[i_1i_2i_3}{}^{j_1}\phi_{i_4]}{}^{j_2j_3j_4}
\aomega_{j_1j_2j_3j_4}&=& (42\aomega^\rep{1}-24\aomega^\rep{7}+
6\aomega^\rep{35})_{i_1i_2i_3i_4}\nn
\phi^{j_1j_2j_3}{}_{i_1}\aomega^\rep{7}_{i_2j_1j_2j_3}&=&
\phi^{j_1j_2j_3}{}_{[i_1}\aomega^\rep{7}_{i_2]j_1j_2j_3} \ . \eea

\makeatletter
\renewcommand{\theequation}{D.\arabic{equation}}
\@addtoreset{equation}{section} \makeatother
\section{Spin Connection Components}

Using the definitions for ${\tilde d}$ and $\CD$ in
({\ref{eqn:notata}}) and ({\ref{eqn:notatb}}), note that \bea de^+
&=& L^{-1} \big( \CD \lambda - (\CD A) \wedge \sigma \big) + e^+
\wedge \big( L^{-1} \CD L+ \partial_u \lambda - \partial_u A
\sigma \big) \nn &+& e^9 \wedge \big[ (LC)^{-1} \big(\partial_z
\lambda -A \partial_u \lambda + (\partial_u A) \lambda -
 {\tilde d}A \big) \big]
\nn &+& LC^{-1} \big( \partial_z L - A \partial_u L + L \partial_u
A \big) e^+ \wedge e^9 \nn d e^- &=& \big({1 \over 2} \CD \cF
\wedge (A \sigma - \lambda)+ \CD \nu - (\CD B)\wedge \sigma \big)
\nn &+& e^+ \wedge \big[ L \big(-{1 \over 2} {\tilde d} \cF +({1
\over 2} \partial_z \cF - \partial_u B) \sigma + \partial_u \nu
\big) \big] \nn &+& e^9 \wedge \big[ C^{-1} \big({1 \over 2} A
{\tilde d} \cF +(\partial_u B -{1 \over 2} \partial_z \cF )
\lambda - {\tilde d} B + \partial_z \nu - A \partial_u \nu \big)
\big] \nn &+& L C^{-1} \big(-{1 \over 2}  \partial_z \cF +
\partial_u B \big) e^+ \wedge e^9 \nn de^9 &=& C \CD \sigma + e^+
\wedge \big( LC \partial_u \sigma \big) +e^9 \wedge \big(-C^{-1}
\CD C + \partial_z \sigma - A \partial_u \sigma \big) \nn &+&
LC^{-1} (\partial_u C) e^+ \wedge e^9 \nn de^i &=& \CD e^i + e^+
\wedge \big(L \partial_u e^i\big) + e^9 \wedge \big[ C^{-1} \big(
\partial_z e^i - A \partial_u e^i \big) \big] \ . \eea

Using these expressions, the following non-vanishing components of
the spin connection are obtained:
\begin{eqnarray}
\omega_{9-+}&=&\frac{C^{-1}}{2}(A\partial_u \log L^{-1}+\partial_u
A-\partial_z \log L^{-1})\nn \omega_{i-+}&=&-\frac{1}{2}(\CD\log
L^{-1} +\partial_u A ~\sigma-\partial_u \lambda)_i
\end{eqnarray}

\begin{eqnarray}
\omega_{+-9}&=&-\frac{C^{-1}}{2}(A\partial_u \log
L^{-1}+\partial_u A-\partial_z \log L^{-1})\nn
\omega_{i-9}&=&-\frac{(L C)^{-1}}{2}(\tilde{d}A +A\partial_u
\lambda-\partial_z \lambda-\partial_u A~ \lambda)_i
\end{eqnarray}

\begin{eqnarray}
\omega_{+-i}&=&\frac{1}{2}(\CD \log L^{-1} +\partial_u A
~\sigma-\partial_u \lambda)_i\nn \omega_{9-i}&=&\frac{(L
C)^{-1}}{2}(\tilde{d}A +A\partial_u \lambda-\partial_z
\lambda-\partial_u A~ \lambda)_i\nn
\omega_{j-i}&=&\frac{L^{-1}}{2}(\CD \lambda-\CD A \wedge
\sigma)_{ij}
\end{eqnarray}

\begin{eqnarray}
\omega_{-+i}&=&\frac{1}{2}(\CD \log L^{-1} +\partial_u A
~\sigma-\partial_u \lambda)_i\nn \omega_{++i}&=&L[(\partial_u
B-\frac{1}{2}\partial_z {\cal F} )\sigma-\partial_u
\nu+\frac{1}{2}\tilde{d}{\cal F})]_i\nn
\omega_{9+i}&=&-\frac{C^{-1}}{2}[(\partial_u B
-\frac{1}{2}\partial_z {\cal F})\lambda +\partial_z
\nu-A\partial_u\nu+\frac{1}{2}A\tilde{d}{\cal F} -\tilde{d}B]_i\nn
&-&\frac{L C}{2}(\partial_u \sigma)_i\nn \omega_{j+i}&=&-L
M_{(ij)}+\frac{1}{2}[\CD \nu-\CD B\wedge \sigma+\frac{1}{2}\CD
{\cal F}\wedge (A\sigma-\lambda)]_{ij}
\end{eqnarray}

\begin{eqnarray}
\omega_{++9}&=&-L C^{-1}(\partial_u B-\frac{1}{2}\partial_z {\cal
F})\nn \omega_{-+9}&=&-\frac{C^{-1}}{2}(A\partial_u \log
L^{-1}+\partial_u A-\partial_z \log L^{-1})\nn
\omega_{9+9}&=&-L\partial_u \log C\nn
\omega_{i+9}&=&\frac{C^{-1}}{2}[(\partial_u B
-\frac{1}{2}\partial_z {\cal F} )\lambda +\partial_z
\nu-A\partial_u\nu+\frac{1}{2}A\tilde{d}{\cal F} -\tilde{d}B]_i\nn
&-&\frac{L C}{2}(\partial_u \sigma)_i
\end{eqnarray}

\begin{eqnarray}
\omega_{-9i}&=&\frac{(L C)^{-1}}{2}(\tilde{d}A +A\partial_u
\lambda-\partial_z \lambda-\partial_u A~ \lambda)_i\nn
\omega_{+9i}&=&-\frac{C^{-1}}{2}[(\partial_u B
-\frac{1}{2}\partial_z {\cal F})\lambda +\partial_z
\nu-A\partial_u\nu+\frac{1}{2}A\tilde{d}{\cal F} -\tilde{d}B]_i\nn
&-&\frac{L C}{2}(\partial_u \sigma)_i\nn \omega_{99i}&=&(\CD \log
C+A\partial_u \sigma-\partial_z \sigma)_i\nn
\omega_{j9i}&=&-C^{-1}\Lambda_{(ij)}+C^{-1}A
M_{(ij)}+\frac{C}{2}(\CD \sigma)_{ij}
\end{eqnarray}

\begin{eqnarray}
\omega_{+ij}&=&-L M_{[ij]}-\frac{1}{2}[\CD \nu-\CD B\wedge
\sigma+\frac{1}{2}\CD {\cal F}\wedge (A\sigma-\lambda)]_{ij}\nn
\omega_{-ij}&=&-\frac{L^{-1}}{2}(\CD \lambda -\CD A \wedge
\sigma)_{ij}\nn \omega_{9ij}&=&-C^{-1}\Lambda_{[ij]}+C^{-1}A
M_{[ij]}-\frac{C}{2}(\CD \sigma)_{ij}\nn \omega_{k
ij}&=&\tilde{\omega}_{k
ij}+\sigma_{[i}\Lambda_{|k|j]}+\sigma_{k}\Lambda_{[ij]}
+\sigma_{[i}\Lambda_{j]k}\nn
&-&(A\sigma-\lambda)_{[i}M_{|k|j]}-(A\sigma-\lambda)_{k}M_{[ij]}
-(A\sigma-\lambda)_{[i}M_{j]k}
\end{eqnarray}
where ${\tilde{\omega}}$ denotes the spin connection of the base
space.


\end{document}